\documentclass[a4paper,10pt,twoside]{cpc-hepnp}

\usepackage{multicol}
\usepackage{graphicx,psfig}
\usepackage{booktabs}
\usepackage{amssymb,bm,mathrsfs,bbm,amscd}
\usepackage[tbtags]{amsmath}
\usepackage{lastpage}

\begin{document}

\fancyhead[co]{\footnotesize T. Teubner et al: Update of $g-2$ of the muon and $\Delta\alpha$}

%\footnotetext[0]{Received 14 January 2010}

\title{Update of $g-2$ of the muon and $\Delta\alpha$}

\author{%
      T. Teubner$^{1}$\email{thomas.teubner@liverpool.ac.uk}%
\quad K. Hagiwara$^{2}$
\quad R. Liao$^{1}$
\quad A.D. Martin$^{3}$
\quad Daisuke Nomura$^{2}$
}
\maketitle

\address{%
1~(Department of Mathematical Sciences, University of Liverpool,
Liverpool L69 3BX, U.K.)\\ 
2~(Theory Center, KEK, Tsukuba, Ibaraki 305-0801, Japan)\\
3~(Department of Physics and Institute for Particle Physics
Phenomenology, University of Durham, Durham DH1 3LE, U.K.)\\
}

\begin{abstract}
We update our Standard Model predictions for $g-2$ of the muon and for
the hadronic contributions to the running of the QED coupling,
$\Delta\alpha^{(5)}_{\rm had}(M_Z^2)$. Particular emphasis is put on
recent changes in the hadronic contributions from new data in the
$2\pi$ channel and from the energy region just below 2 GeV. 
\end{abstract}



\begin{multicols}{2}

\section{Introduction}
\vspace{-3mm}
The anomalous magnetic moment of the muon, $a_{\mu}=(g-2)/2$, has been
the subject of wide interest and detailed research. The discrepancy between its
experimental value as measured by BNL \cite{BNL} and its prediction
within the Standard Model (SM) is one of the few -- if not the only --
experimental sign of physics beyond the SM (apart from neutrino
mixing). This has triggered a lot of intense scrutiny of both the
experimental determination and the theoretical evaluation of
$a_{\mu}$. The BNL experiment E821 has achieved an impressive precision of
0.5ppm \cite{BNL}, and further improvements may only be reached with
the planned experiments at Fermilab and J-PARC, see the presentations
\cite{LRoberts,TMibe}. As discussed below, the SM prediction relies
heavily on the experimental information of the measured hadronic cross
sections at low energies. During the last 15 years, the SM prediction
has gone through several phases of improvement and consolidation and
has now, for the first time, reached an accuracy even slightly better
than the experimental value. This is mainly due to the big efforts to
measure the hadronic cross sections with increasing accuracy and the
progress of various groups working on the data-driven evaluation of
the hadronic contributions, which are in fairly good agreement (though
further improvements are foreseen as will be discussed briefly in
Section~\ref{secoutlook}). In this article we will concentrate on the
main changes in the hadronic contributions to $a_{\mu}$, updating the
works \cite{HMNT06,HMNT03}. 

Similarly to $g-2$, the theoretical uncertainties in the running of the QED
coupling, $\alpha(q^2)$, are completely dominated by the hadronic
contributions, $\Delta\alpha^{(5)}_{\rm had}(q^2)$. Recall 
$\alpha(M_Z^2)$ is the least well known of the set of precision
observables $[G_F, M_Z, \alpha(M_Z^2)]$, so its error is a limiting
factor in the electroweak fits of the SM as performed e.g. by the LEP
Electroweak Working Group.

\section{Standard Model prediction of  $g-2$}
\vspace{-3mm}
The anomalous magnetic moment of the muon receives
contributions from all sectors of the SM. The
QED and electroweak (EW) corrections can be calculated within
perturbation theory and are, through many impressive works, well
under control. 
In the compilation for $a_{\mu}^{\rm SM}$ presented here we use 
$a_{\mu}^{\rm QED} = 116584718.08(15) \cdot 10^{-11}$
\cite{Kinoshitaetal,Aoyamaetal} and $a_{\mu}^{\rm EW} = (154 \pm 2)\cdot
10^{-11}$ \cite{Czarneckietal}, as e.g. reviewed in \cite{Passera}. 
The hadronic contributions can not be reliably calculated in
perturbative QCD (pQCD) as the loop-integrals are dominated by low
momentum transfer, i.e. the non-perturbative region of QCD. They are
typically divided into the leading (LO) and higher-order (HO) vacuum
polarisation (VP), and the so-called light-by-light contributions,
which are also subleading: 
$a_{\mu}^{\rm had} = a_{\mu}^{\rm had,\,LO\,VP} + a_{\mu}^{\rm had,\,HO\,
  VP} + a_{\mu}^{\rm had,\,l-by-l}$. While the VP induced corrections
can be calculated with methods based on dispersion relations and using
experimentally measured hadronic cross sections as input, the
light-by-light scattering contributions can be estimated only using 
models.\footnote{First principle calculations within lattice gauge
  field theory are underway but very difficult and at an early stage.}
The results from different groups vary considerably, both w.r.t. the mean
value and the error. For a review presented at this conference see
\cite{Nyffeler}. In the SM prediction of $g-2$ presented here we use
the value $a_{\mu}^{\rm had,\,l-by-l} = (10.5 \pm 2.6)\cdot 10^{-10}$,
which has been obtained in \cite{Pradesetal} as a combination of
results based on different models.\footnote{Note that this is slightly
  different from (though certainly compatible with) the value $a_{\mu}^{\rm
    had,\,l-by-l} = (116 \pm 40)\cdot 10^{-11}$ as obtained in
  \cite{Nyffelerpub,JegerlehnerNyffeler} and discussed in
  \cite{Nyffeler}.} In the following we will discuss in more detail
recent changes in the VP contributions.

\subsection{Hadronic VP contributions}
The LO hadronic VP contributions are calculated using the dispersion integral
\begin{equation}
a_{\mu}^{\rm had,\,LO\,VP} = \frac{1}{4\pi^3} \int_{m_{\pi}^2}^{\infty} {\rm d}s 
\, \sigma_{\rm had}^0(s) K(s)\,,
\label{eq:disp}
\end{equation}
where $K(s) = \frac{m_{\mu}^2}{3s} \cdot (0.4 \ldots 1)$ is a known
kernel function giving highest weight to lowest energies
$\sqrt{s}$, and $\sigma_{\rm had}^0(s)$ is the hadronic cross
section for $e^+ e^- \to \gamma^* \to hadrons\,(+\gamma)$. The superscript
$^0$ indicates that the `undressed' cross section must be used, i.e. the
cross section without VP effects in the virtual photon, but including
final state radiation (FSR) of photons. To arrive at the best compilation for
$\sigma_{\rm had}$, at low energies ($\sqrt{s}<2$ GeV) about 24
hadronic channels (exclusive final states) have to be summed, and in
each channel the data from different experiments have to be
combined. At intermediate energies $\sigma_{\rm had}$ is measured
inclusively. Perturbative QCD can only be used away from resonances, and
(most) data driven analyses use pQCD only for energies above the open
$b\bar b$ threshold. For details of the data input, their treatment
w.r.t. radiative corrections and the data combination through a
non-linear $\chi^2_{\rm min}$ fit, as used in the
work reported here, see \cite{HMNT06,HMNT03}. Note that there are
uncertainties w.r.t. the correct application of radiative corrections
(undressing of VP, possible addition of neglected photon FSR) to the
data, especially in the case of older data sets. In our analysis this
leads to the assignment of a separate error due to radiative
corrections, $\delta a_{\mu}^{\rm had,\,VP + FSR} \simeq 1.8 \cdot
10^{-10}$, which alone is about ten times bigger than the uncertainty
of the electroweak contributions $a_{\mu}^{\rm EW}$. 

\subsubsection{Recent changes in the $2\pi$ channel}
More than 70\% of $a_{\mu}^{\rm had}$ is coming from the $\rho \to 2\pi$
channel. Following older experiments, the cross section $e^+ e^- \to \pi^+\pi^-$
has been measured in recent years with increasing accuracy by the Novosibirsk
experiments CMD-2 and SND, see e.g. \cite{Ignatov} for a short review of
their results. Figure~\ref{fig1} displays the impressive agreement of
the most recent data sets from CMD-2 and SND, together with recent data from
KLOE \cite{KLOEpub} obtained with the method of Radiative Return (see
\cite{MCSIGHAD} for a detailed review of the method and its
application, and \cite{KLOE} for the very new KLOE analysis of the
$2\pi$ channel presented at the PhiPsi09 conference but not yet
available for public use). The grey band shows the result of our data
combination in this channel including also older data, but excluding
the KLOE data, see the discussion below. 
\begin{center}
\vspace{1mm}
\psfig{file=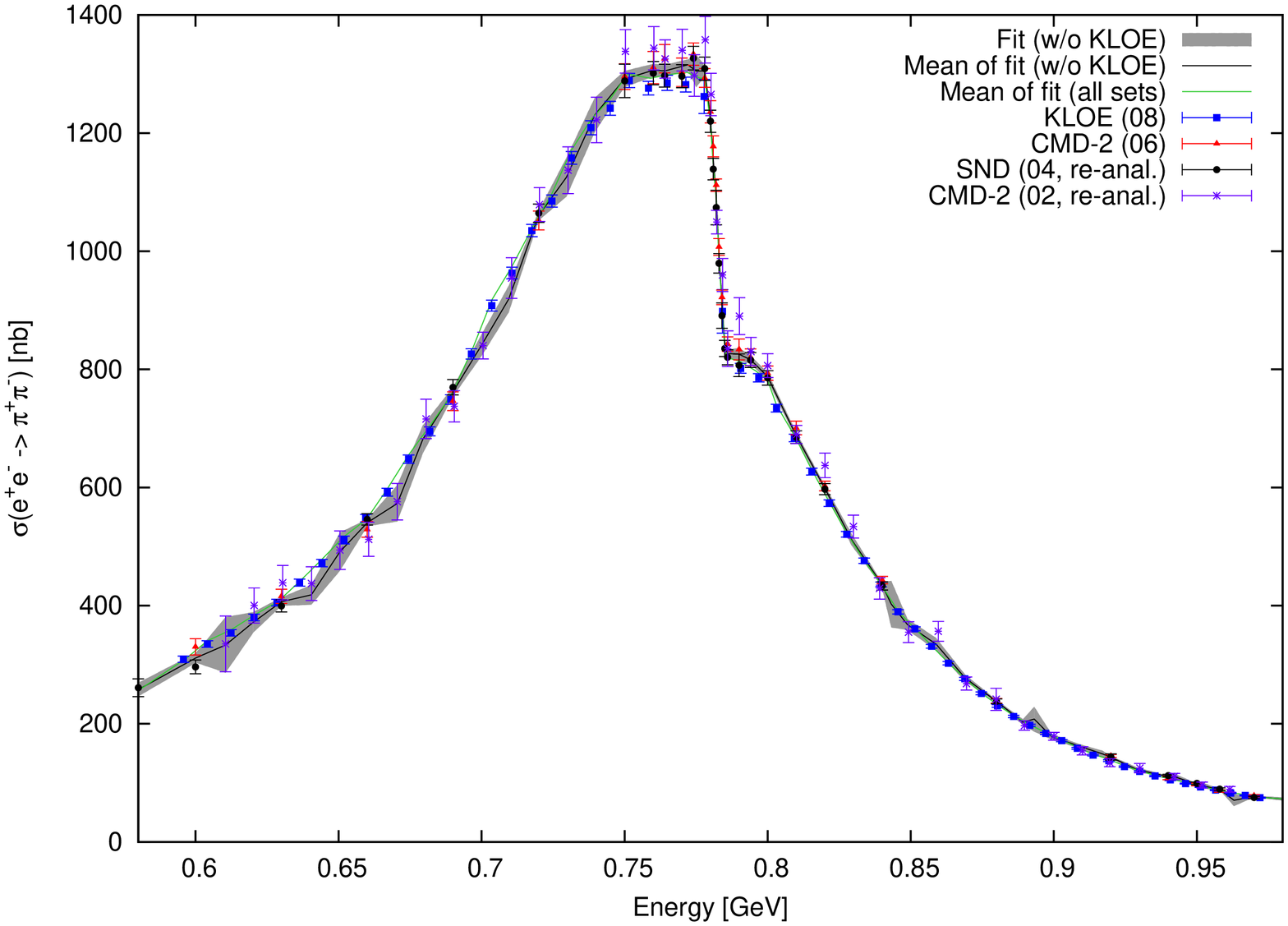,width=8cm,angle=270}
\vspace{1mm}
\figcaption{\label{fig1} Most important data in the $2\pi$ channel and
  fits as described in the text.}
\end{center}
\begin{center}
\vspace{1mm}
\psfig{file=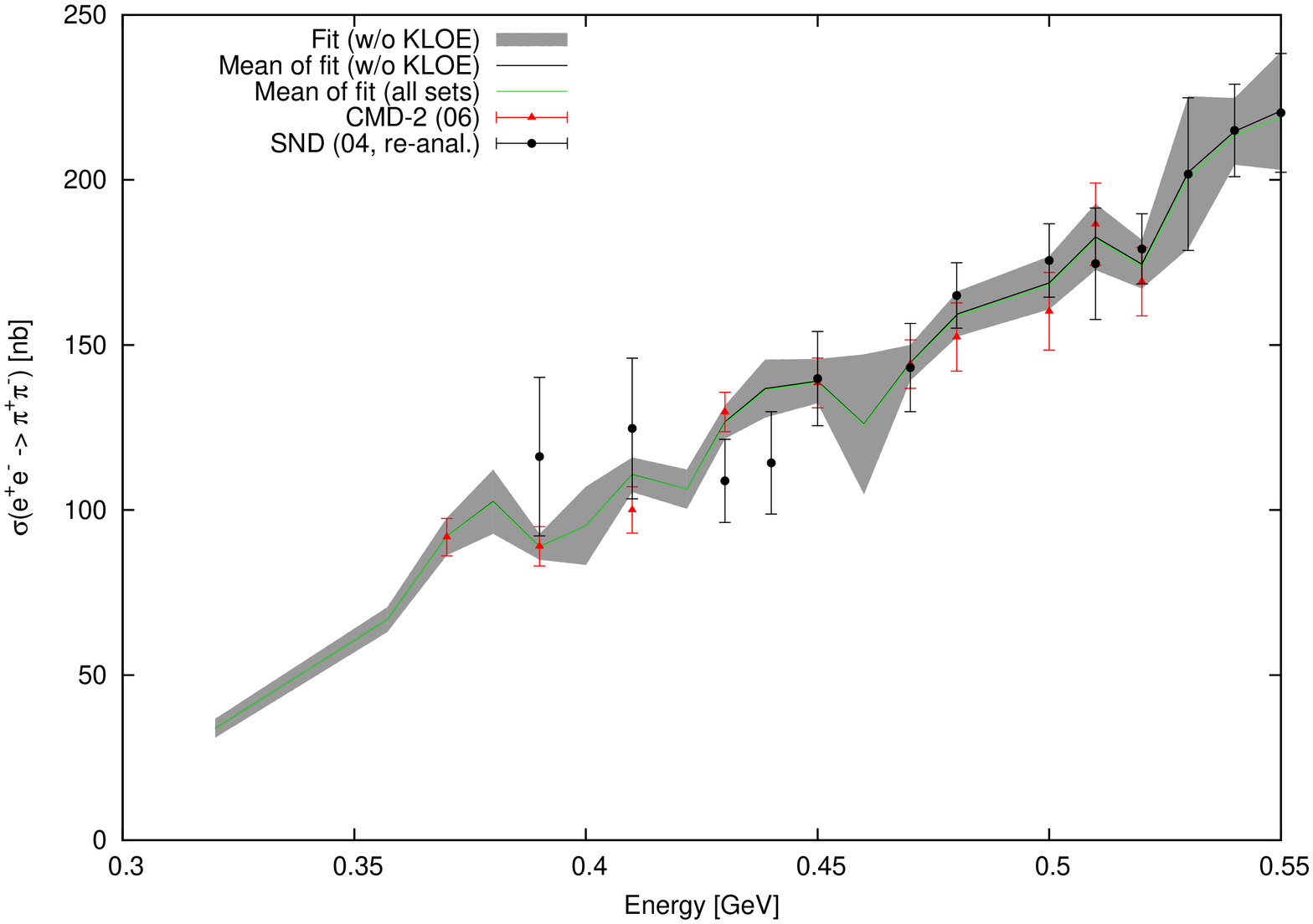,width=8cm,angle=270}
%\vspace{1mm}
\figcaption{\label{fig2} Low energy region close to the $2\pi$ threshold.} 
\end{center}
\vspace{-2mm}
The lowest energy region close to the $2\pi$ threshold is displayed in
Fig.~\ref{fig2}. The recent CMD-2 data represented by (red) triangles
and error bars clearly demonstrate the improvement from this single
data set alone, with the fit in this region being dominated by these
data. Figure~\ref{fig3} shows an enlargement of the $\rho$-$\omega$
interference region which is now very well mapped out by the
consistent data sets. However, Fig.~\ref{fig3} also shows that the
KLOE data are undershooting the combination of the other data in this
region, while typically being higher than other data at lower energies
as can be seen by the (green) solid line in Fig.~\ref{fig1}. 
This apparent difference in shape is highlighted in Fig.~\ref{fig4},
which shows the normalised difference of the KLOE cross section and
the combination of the other $2\pi$ data by the square markers, while
the band displays the size of the error of the compilation without KLOE.
(The bands displayed in the figures are obtained from the diagonal
elements of the fit's full covariance matrix.) 
\begin{center}
\psfig{file=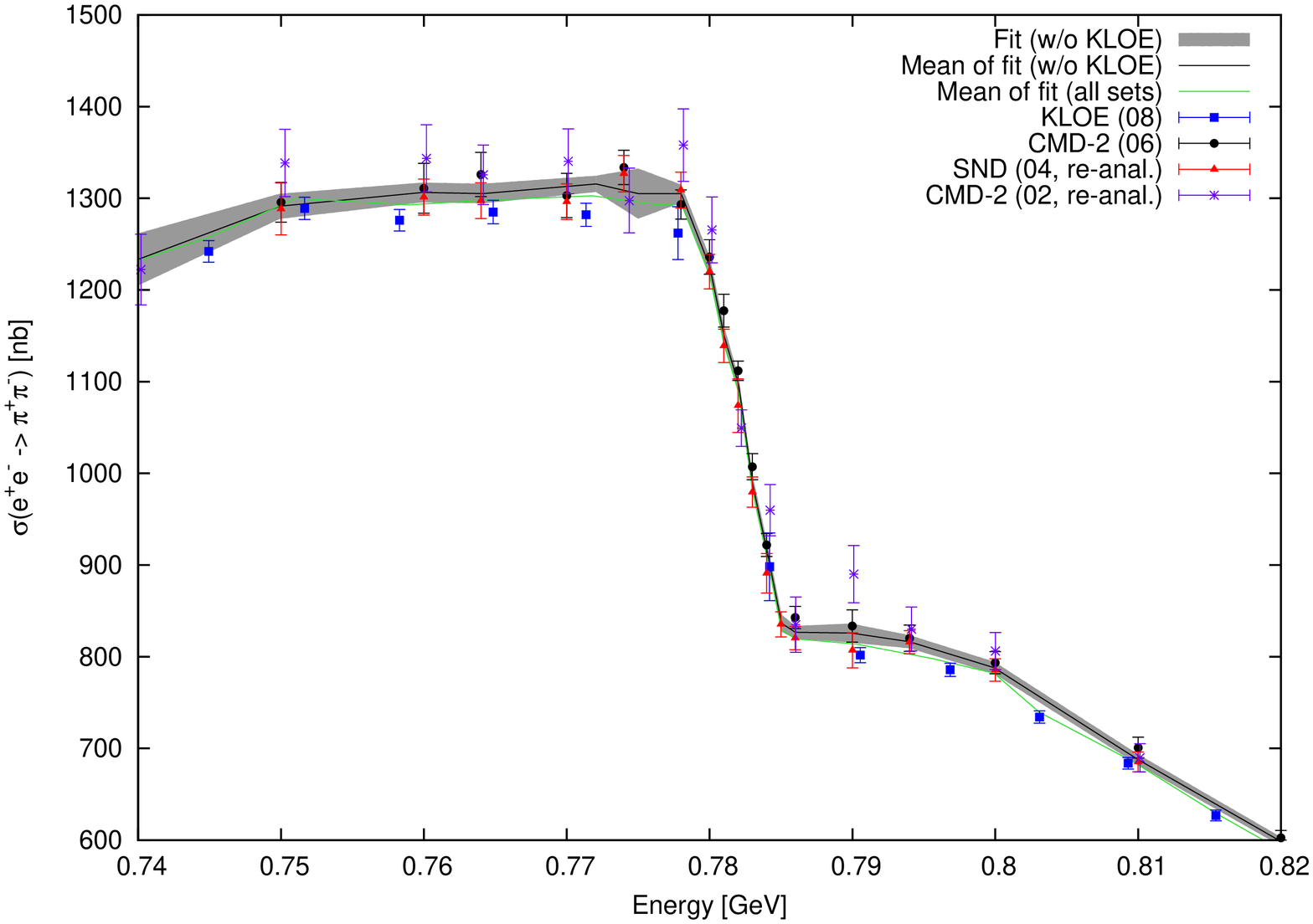,width=8cm,angle=270}
\vspace{-2mm}
\figcaption{\label{fig3} $\rho$-$\omega$ interference region in the
  $2\pi$ channel.} 
\end{center}
\begin{center}
\vspace{-3mm}
\psfig{file=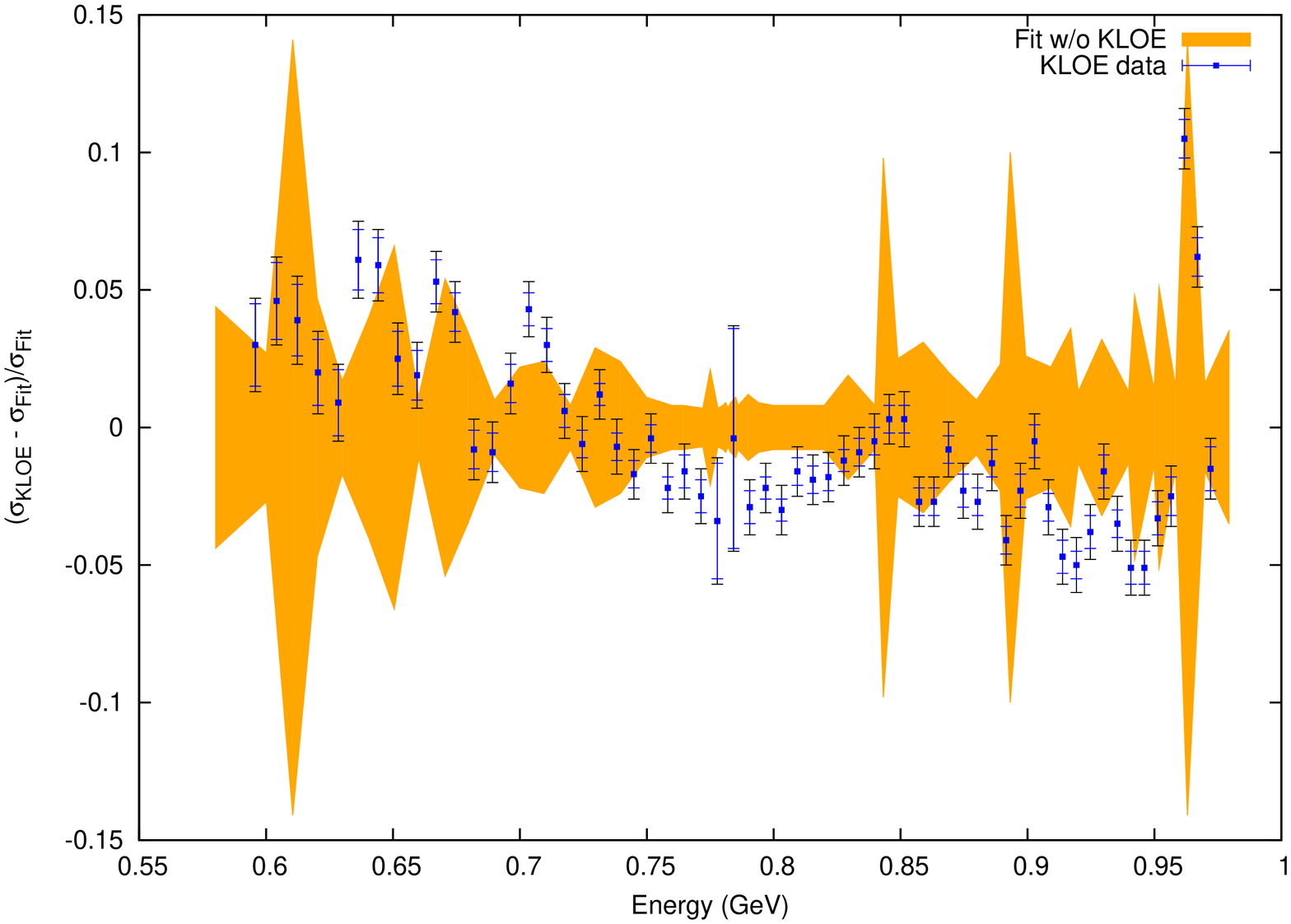,width=8cm,angle=270}
\vspace{1mm}
\figcaption{\label{fig4} Normalised difference of the KLOE data
  \cite{KLOEpub} and the data compilation excluding KLOE; the band
  represents the error of the compilation.} 
\end{center}

One should keep in mind that the KLOE data are obtained via the method
of Radiative Return at fixed collider centre-of-mass energy, whereas
the other data are measured via the traditional method of energy scan
by adjusting the $e^+e^-$ beam energies. Hence Monte Carlo simulation 
tools including  radiative corrections and also the systematic effects
are completely different between the two approaches. Up to now it is
not clear what causes the difference in the shapes of the $2\pi$ data,
and more studies are underway to clarify the situation. Unfortunately
this difference in shape prevents us from including the KLOE data in a
straightforward way in the non-linear $\chi^2_{\min}$
fit.\footnote{The fit allows a readjustment of the overall
  normalisation of the data sets within their systematic
  errors. Including the KLOE data would lead to a bad $\chi^2_{\rm
    min}/{\rm d.o.f.}$ and unnatural normalisation effects pulling
  the fit upward, see \cite{HMNT06} for a detailed discussion.} 
Note that if we calculate the contribution to $g-2$ from the KLOE data
alone, we obtain $a_{\mu}^{\pi\pi,\, {\rm KLOE}} = (384.16 \pm 3.47)\cdot
10^{-10}$, in perfect agreement with the result of the integral over our 
compilation of all $2\pi$ data without KLOE (but in the range of the
KLOE data), for which we get $a_{\mu}^{\pi\pi,\,{\rm w/out\ KLOE}} =
(384.12 \pm 2.51)\cdot 10^{-10}$. 
We therefore use the procedure adopted already in \cite{HMNT06} where
earlier KLOE data (now superseded by \cite{KLOEpub}) were included
{\em after} integration, by calculating a weighted average with the integral
over the data compilation without KLOE, and not performing a
point-by-point combination.

BaBar has also published their first measurement of the $2\pi$ channel
based on Radiative Return \cite{Babar2pi} (see also
\cite{WangBabar2pi}), finding some discrepancies with KLOE. We have
not included the new BaBar data in the analysis presented here as they
were not available for public use at the time of the conference, but
see \cite{Davier2pi,Davier} for an analysis which includes them.

\subsubsection{Energy region below 2 GeV}
Important changes in the data input have also happened in the region
between 1.4 and 2 GeV. This region is particularly difficult, as a
growing number of multi-hadron exclusive final states becomes
accessible and has to be included to obtain $\sigma_{\rm had}$ with
good accuracy.  However, these energies are above the reach of the
Novosibirsk machine\footnote{This will change with the current
  upgrade, see \cite{CMD3}.}, and the quality of the available data
was not very good. Alternatively, one can rely on inclusive $R$
measurements, but for this only rather old and not very precise data
are available. This situation has changed with BaBar measuring,
through Radiative Return, many channels with higher accuracy than
earlier experiments. These include new data for $2\pi^+2\pi^-$
\cite{Babar4pi}, $K^+K^-\pi^0$, $K^0_S\pi K$ \cite{BaBar_1},
$2\pi^+2\pi^-\pi^0$, $K^+K^-\pi^+\pi^-\pi^0$, $2\pi^+2\pi^-\eta$
\cite{BaBar_2}, $2\pi^+2\pi^-2\pi^0$ \cite{BaBar_3} used for our
updated analysis. 
Figures~\ref{fig5} -- \ref{fig7} exemplify the influence of the new
BaBar data. The new data are not always in good agreement with other
sets; in such cases the fit has a poor quality and we scale up the
error of the channel's contribution by $\sqrt{\chi^2_{\rm min}/{\rm
    d.o.f.}}$ (e.g. in the $2\pi^+ 2\pi^- 2\pi^0$ channel $\chi^2_{\rm
  min}/{\rm d.o.f.} = 2.7$.) 

\subsubsection{Sum rule analysis}
Note that in previous $g-2$ analyses \cite{HMNT03,HMNT06} we 
found a discrepancy between using the available inclusive data in the
region 1.4 to 2 GeV and adding the exclusive channels. The inclusive data
were lower than the sum of the exclusive channels, though they were
found to be similar in shape. 
\begin{center}
\vspace{1mm}
\psfig{file=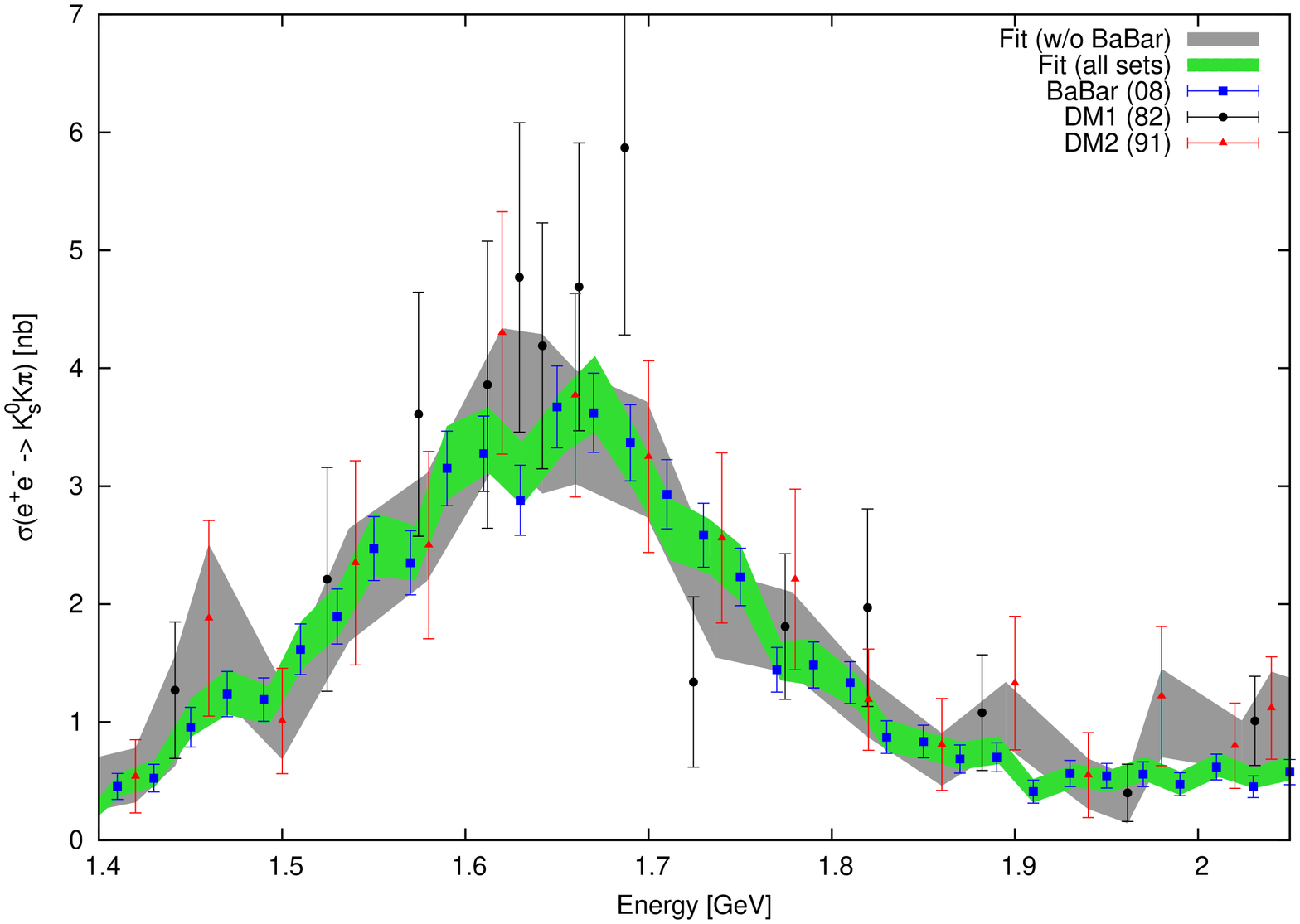,width=8cm,angle=270}
\vspace{1mm}
\figcaption{\label{fig5} $K_S^0 K \pi$ channel with improvement due to
  recent BaBar data \cite{BaBar_1}.} 
\end{center}
\begin{center}
\psfig{file=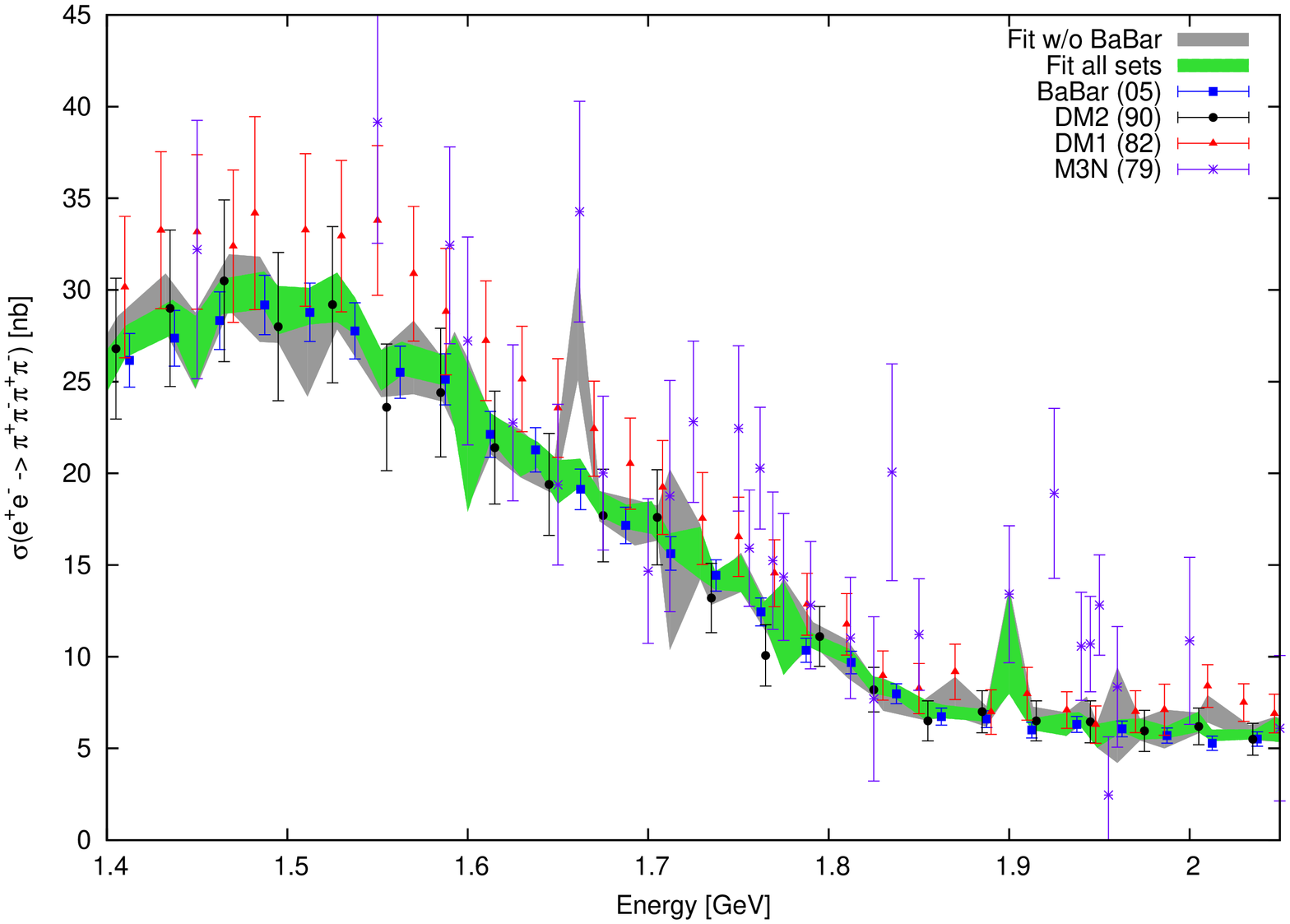,width=8cm,angle=270}
\vspace{-1mm}
\figcaption{\label{fig6} $2\pi^+ 2\pi^-$ channel.} 
\end{center}
\begin{center}
\vspace{-3mm}
\psfig{file=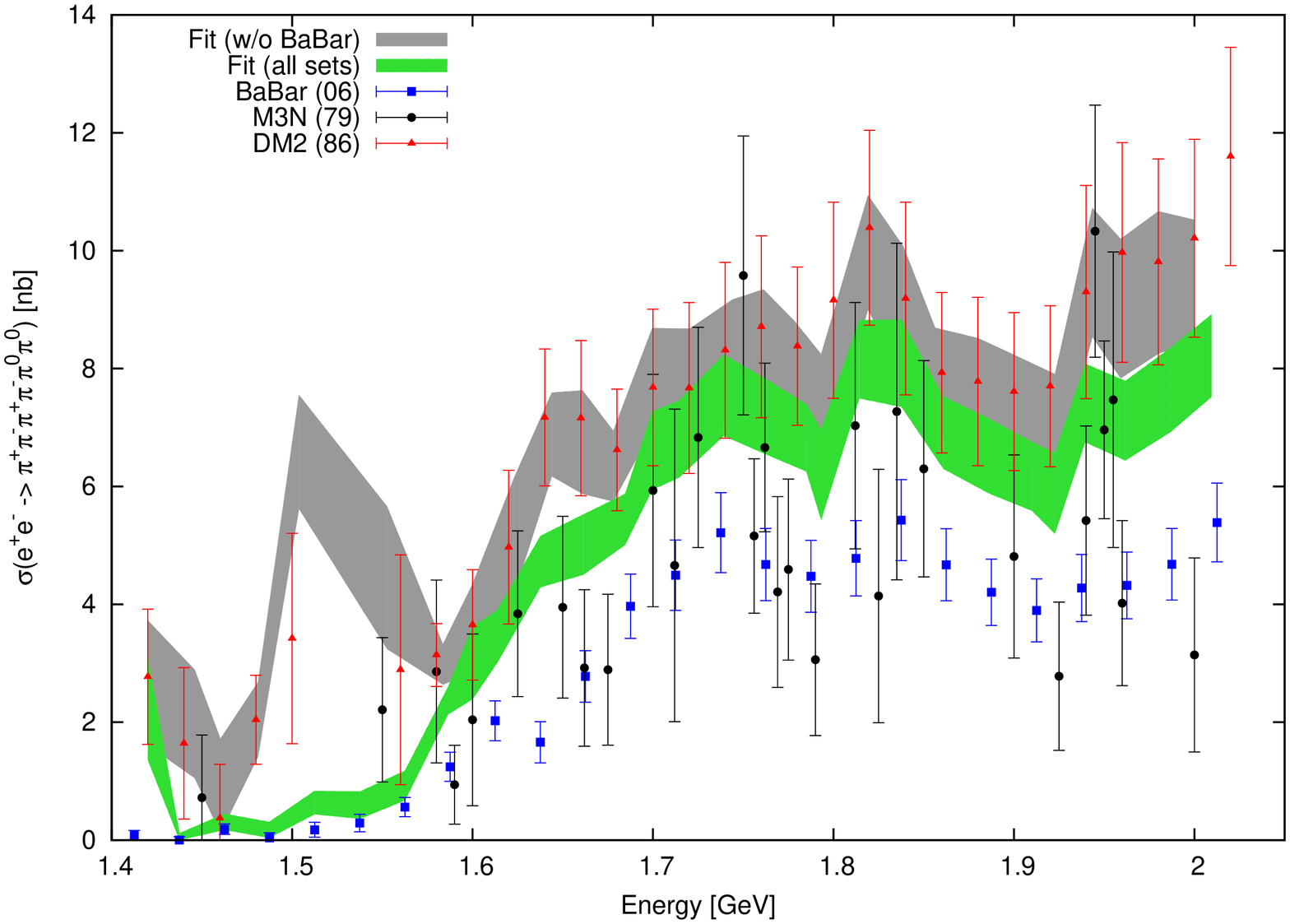,width=8cm,angle=270}
\vspace{-1mm}
\figcaption{\label{fig7} $2\pi^+ 2\pi^- 2\pi^0$ channel.} 
\vspace{-3mm}
\end{center}
In \cite{HMNT03} we performed a QCD sum
rule analysis, concluding that the inclusive data are more 
compatible with pQCD and the world average of $\alpha_s$, and
therefore chose to use the inclusive data instead of the sum over the
exclusive for energies $\sqrt{s}=1.43 \ldots 2$ GeV.
Since then the situation has changed: the hadronic data has changed
slightly, being lower at low energies and also at energies above 2
GeV. Also, with the inclusion of the recent BaBar data the sum of the
exclusive channels in the region 1.4 to 2 GeV has become slightly
lower and more accurate than before. Our updated sum rule analysis is
summarised in Fig.~\ref{fig8}; different sum rules based on pQCD are made to
match the corresponding sum rule integrals over the data by fitting
for $\alpha_s$ as a free parameter (see \cite{HMNT03} for details). It
is clear that the new sum over exclusive channels is more accurate
than the old inclusive data and also more compatible with the
predictions based on pQCD with a world average value of $\alpha_s$.
We therefore are now combining the results from the inclusive and the
sum over exclusive data in this energy region.\footnote{Note that the
  sum over exclusive channels stills requires, due to the lack of
  experimental information, the use of iso-spin relations for unknown
  channels, which in turn results in a large error from the poorly
  known $K\bar K \pi\pi$ channel.} 
\begin{center}
\vspace{-5mm}
\psfig{file=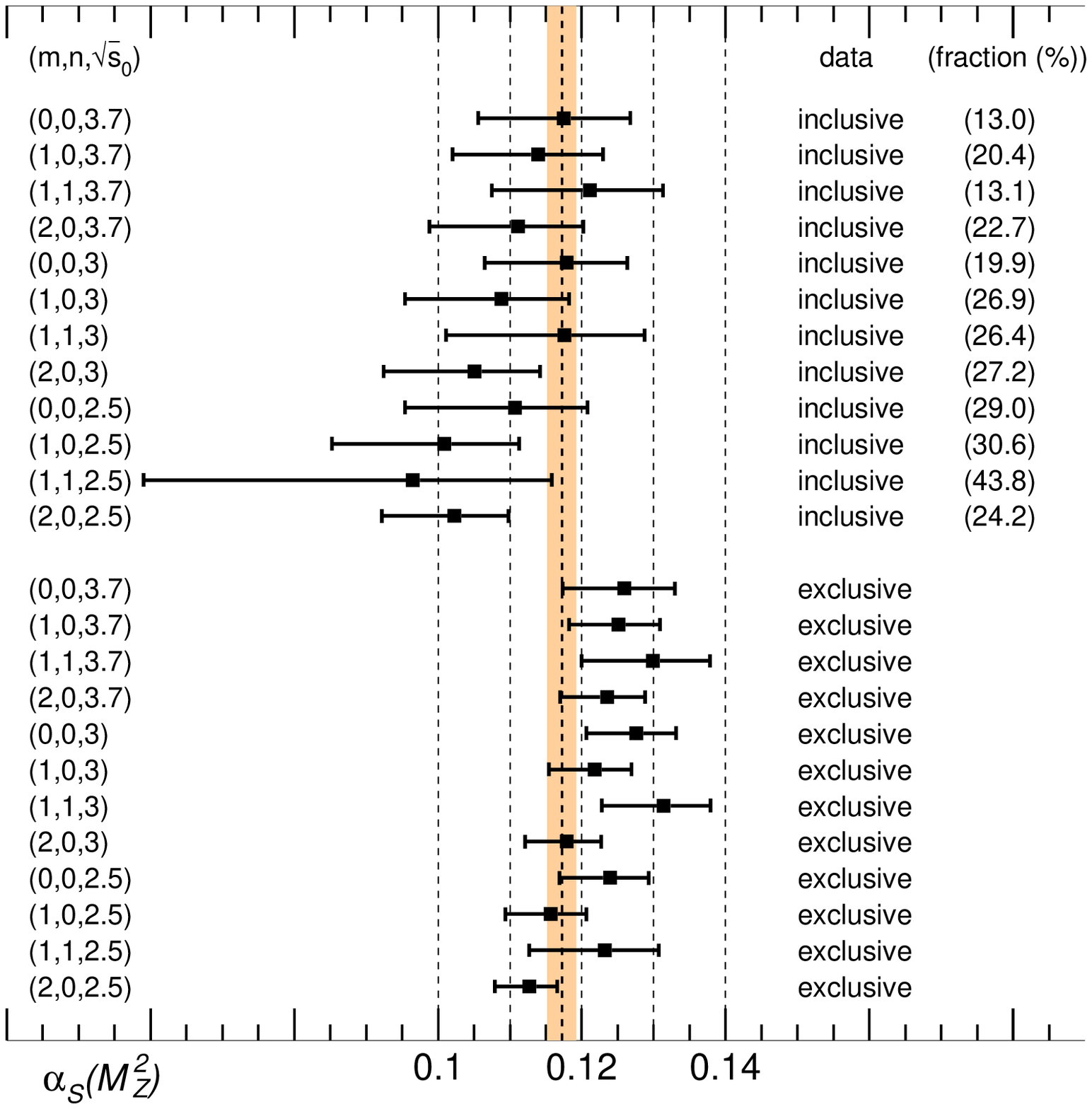,width=9cm,angle=0}
\vspace{-5mm}
\figcaption{\label{fig8} Results for different sum-rules \cite{HMNT03}
  translated into a prediction of $\alpha_s$; the band shows the world
  average of $\alpha_s(M_Z^2)$.} 
\end{center}

\subsubsection{Other changes and result for $a_{\mu}^{\rm had,\,VP}$}
In addition to the important changes discussed above, compared to
\cite{HMNT06}, we have also included new data in other channels: 
$K^+K^-$ from CMD-2 \cite{CMD2_1} and SND \cite{SND_1}, $K^0_S K^0_L$ from
SND \cite{SND_2}, $\pi^+\pi^-\pi^0$ from CMD-2 \cite{CMD2_2},
$\omega\pi^0$ from KLOE \cite{KLOEomegapi}, and inclusive $R$ data at
higher energies above 2 GeV from BES \cite{BES_1,BES_2} and CLEO
\cite{CLEO}. Figure~\ref{fig9} shows the recent BES data together with
the fit of all inclusive data in this region and the prediction from
pQCD. While the contribution to $g-2$ is significantly smaller than
previously, it is obvious that pQCD, which is in perfect agreement
with the three latest BESII (09) data points \cite{BES_2}, is still
somewhat lower than the fit for energies below 3 GeV. 
\begin{center}
\vspace{1mm}
\psfig{file=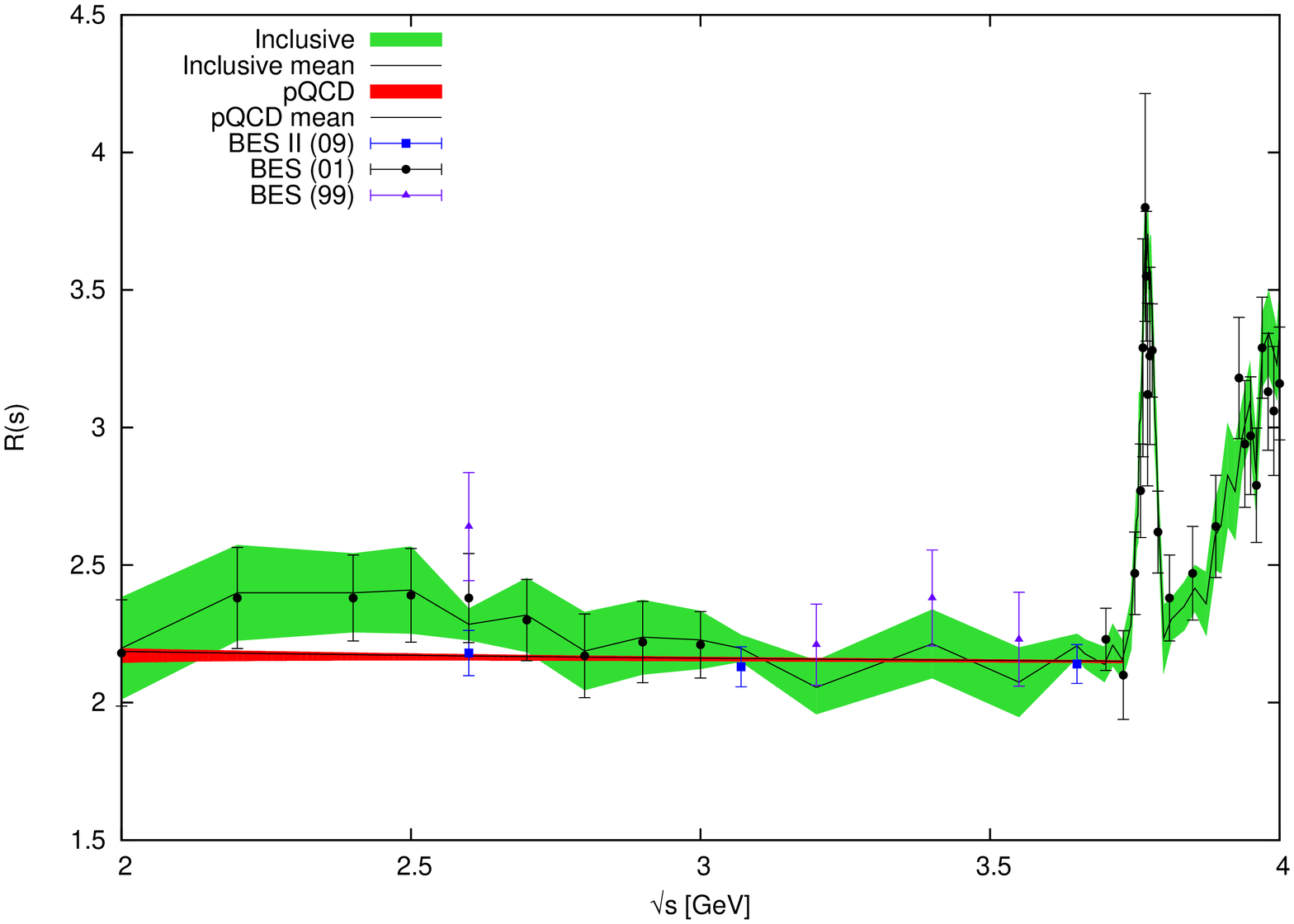,width=8cm,angle=270}
\vspace{1mm}
\figcaption{\label{fig9} Recent data from BES together with the fit of
  all inclusive data compared to pQCD in the energy region above 2 GeV.} 
\end{center}

Numerically the changes to $a_{\mu}^{\rm had,\,LO\,VP}$ from the
different energy regions amount to (units of $10^{-10}$, compared to
results from \cite{HMNT06}): $-0.76$ ($0.32 - 1.43$ GeV, low energy
exclusive channels), $+2.10$ ($1.43 - 2$ GeV, results from inclusive
and sum over exclusive data combined), $-1.35$ ($2 - 11.09$ GeV,
higher energy inclusive data). This accidentally leads to a near
perfect cancellation of the shifts, with the total result
$a_{\mu}^{\rm had,\,LO\,VP} = (689.4 \pm 3.6_{\rm exp} \pm 1.8_{\rm
  rad})\cdot 10^{-10}$. The first error is coming from the statistical
and systematic error of the data, whereas the second error is our
estimate of the uncertainty in the radiative corrections as mentioned
above. However, compared to our earlier result from \cite{HMNT06},
$a_{\mu}^{\rm had,\,LO\,VP}({\rm HMNT06}) = (689.4 \pm 4.2_{\rm exp}
\pm 1.8_{\rm rad})\cdot 10^{-10}$, there is a further reduction in the error. 

The higher-order VP induced contributions can also be calculated using
a dispersion integral, see \cite{HMNT03} for details. Our updated
result is nearly unchanged and reads $a_{\mu}^{\rm had,\,HO\,VP} =
(-9.79 \pm 0.06_{\rm exp} \pm 0.03_{\rm rad})\cdot 10^{-10}$.

\subsection{SM predictions compared to the BNL measurement}
Combining the QED, EW and hadronic contributions as discussed above,
we arrive at our SM prediction of the anomalous magnetic moment of the
muon, 
\begin{equation}
\label{eqamuSM}
a_{\mu}^{\rm SM}({\rm HLMNT09}) = (116\ 591\ 773 \pm 48) \cdot 10^{-11}\,. 
\end{equation}
Due to a small shift of Codata's muon to proton magnetic ratio
$\lambda$ the experimental value for $a_{\mu}$ from BNL is now
$a_{\mu}^{\rm EXP} = 116\ 592\ 089(63) \cdot 10^{-11}$
\cite{LRoberts}. This results 
in a difference $a_{\mu}^{\rm EXP} - a_{\mu}^{\rm SM} = (31.6 \pm
7.9)\cdot 10^{-10}$ which corresponds to a $4\,\sigma$ discrepancy. 
The situation is displayed in Fig.~\ref{fig10} where we also show our
previous result and the most recent results from Jegerlehner and
Nyffeler \cite{JegerlehnerNyffeler} and three different predictions
from Davier et al.~\cite{Davier,Davier2pi,Daviertau}: While their
$e^+e^-$ based result without the new BaBar $2\pi$ (labelled `w/o
BaBar (09)' in the figure) data agrees very well with our evaluation,
the result including these new data lead to a shift upwards, but
still compatible with the other $e^+e^-$ based result. The third
result employs, in addition to $e^+e^-$ data, also the use of $\tau$
spectral function data from ALEPH, OPAL, CLEO and BELLE.
\begin{center}
\vspace{-3mm}
\psfig{file=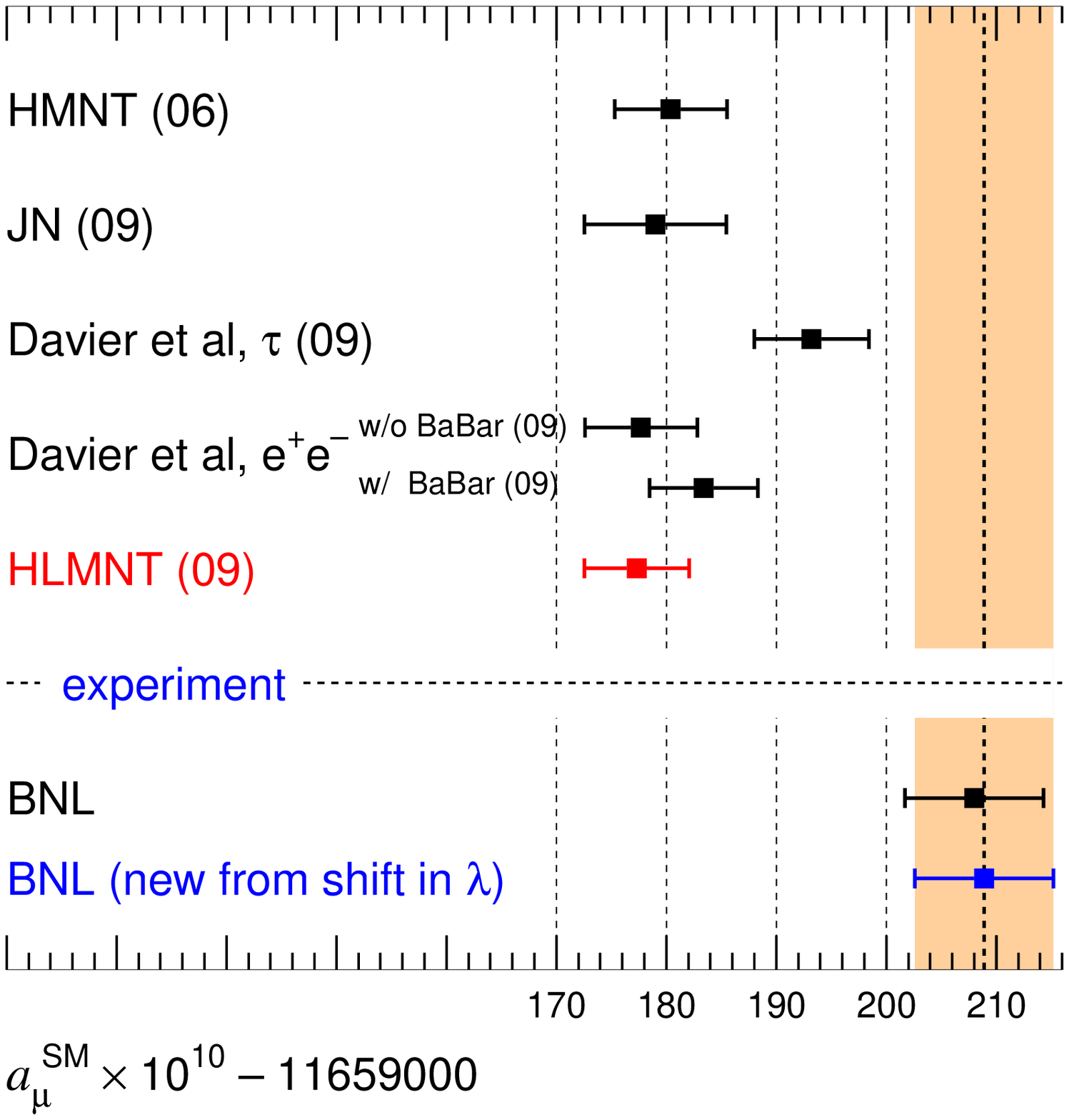,width=7.5cm,angle=0}
\vspace{-2mm}
\figcaption{\label{fig10} Comparison of recent predictions for $g-2$
  compared to the BNL measurement.} 
\end{center}
To translate the charged current induced hadronic $\tau$ decay data
into the required spectral functions requires the application of
isospin breaking corrections, which can only be predicted in a model
dependent way. Earlier $\tau$ based results from Davier
et al.\ were incompatible with $e^+e^-$ based results, but with a
re-evaluation of the isospin breaking corrections
\cite{Daviertau,Lopez,Davier} they find the result displayed in
Fig.~\ref{fig10} (labelled `$\tau$ (09)') which is now marginally
consistent. These findings were discussed controversially at the
PhiPsi09 conference; Benayoun presented an alternative approach based
on Hidden Local Symmetry and dynamical $(\rho, \omega, \phi)$
mixing. With this and with a global fit he gets consistency of the
$\tau$ spectral function with the $e^+e^-$ data and an improved $2\pi$
contribution to $g-2$~\cite{Benayoun}. Keeping in mind that other
studies obtained a very large uncertainty of the isospin breaking
corrections (see e.g. \cite{MelnikovVainshtein}) we believe that,
until this issue is better understood, the predictions of $g-2$ based
on $e^+e^-$ data alone are more reliable.

\section{Running QED coupling $\alpha(q^2)$}
\vspace{-3mm}
Based on the same data compilation, we also predict the hadronic
contributions to the running of the QED coupling,
$\alpha(q^2)=\alpha/(1-\Delta\alpha_{\rm lep}-\Delta\alpha_{\rm
  had})$. The hadronic VP is important for many studies where high
accuracy radiative corrections are required, as is the case for the
hadronic contributions to $g-2$, see \cite{MCSIGHAD} for a very recent
review. Of particular importance as input in EW precision fits is the
quantity $\Delta\alpha_{\rm had}^{(5)}(M_Z^2)$, the hadronic
contributions to the running $\alpha$ from five-flavours (the
contribution from the top quark is usually added using pQCD).  
Our updated value is $\Delta\alpha_{\rm had}^{(5)}(M_Z^2) = 0.02760
\pm 0.00015$. This is slightly higher and significantly more accurate
than the number from Burkhardt and Pietrzyk used as default by
the LEP EW Working Group, and leads e.g. to a lower preferred Higgs
mass and a lower upper limit from the famous `Blue Band Plot'.

\section{Summary and outlook}
\label{secoutlook}
\vspace{-3mm}
We have given an update of the SM prediction of $g-2$, emphasising
recent developments for the hadronic contributions. With the $e^+e^-$
based analysis presented here we obtain a $4\,\sigma$ discrepancy
between the experimentally measured value of $a_{\mu}$ and its SM
prediction. Slightly less significant discrepancies are reported by
other groups, depending on the data used and the details of the
analyses. However, for all $e^+e^-$ based analyses the discrepancy is
about $3-4\,\sigma$ and standing all scrutiny. 

For the future, further improvements of the SM prediction will be
possible. The method of Radiative Return has proven extremely powerful
and will be leading to many more results. In addition to the already
reported new $2\pi$ data from KLOE \cite{KLOE}, further analyses are
ongoing, and there are exciting prospects with KLOE2 \cite{Venanzoni}. 
There is also a rich programme going on at BELLE, and the possibility
of SuperBELLE at the horizon. 
CMD-3 and SND at VEPP-2000 currently commissioned in Novosibirsk are
aiming at largely improving the exclusive measurements in the region
below 2 GeV, whereas BESIII at BEPCII will cover higher energies. 
With all these developments the error of the SM prediction of $g-2$ is
expected to shrink even further, so that in the future the
light-by-light contributions may eventually become the limiting factor. 
Obviously there is an extremely strong case for new the $g-2$
experiments as planned by the $g-2$ Collaboration for Fermilab
\cite{LRoberts} and at J-PARC \cite{TMibe}. With this, $g-2$ will become
even more powerful in establishing and  constraining physics beyond
the Standard Model. 

%
%\begin{center}
%\vspace{-2mm}
%\psfig{file=pie_phipsi09.eps,width=8cm,angle=0}
%\vspace{-2mm}
%\figcaption{\label{fig12} Pie diagrams for contributions and errors to
%  $g-2$ and $\Delta\alpha(M_Z^2)$.} 
%\vspace{-2mm}
%\end{center}
%
%\acknowledgments{We thank  $\cdots$.}

\end{multicols}

\vspace{-2mm}
\centerline{\rule{80mm}{0.1pt}}
\vspace{1mm}

\begin{multicols}{2}

\end{multicols}

\clearpage

\end{document}